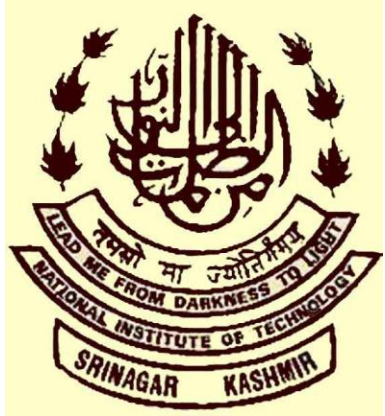

Project Report

On

## Analysis and Design of VANET Protocols for Srinagar City

Submitted in partial fulfilment of the requirements

for the award of the degree of

BACHELOR OF TECHONOLOGY

IN

INFORMATION TECHNOLOGY

FURQAN YAQUB KHAN

IT/03/15

Under the supervision of

DR. Shabir Ahmad Sofi

Department of Information Technology

National Institute of Technology Srinagar,

J&K

June 2019

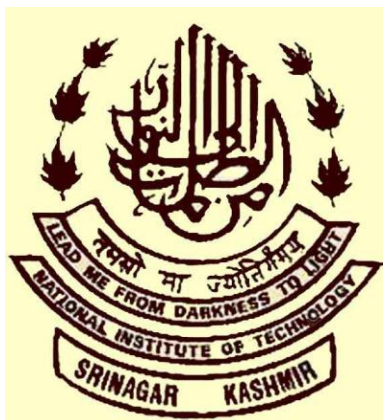

# CERTIFICATE

This is to certify that the project titled Analysis and Design of VANET Protocols for Srinagar City has been completed by Furqan Yaqub Khan (IT/03/15) under my supervision in partial fulfilment of the requirements for the award of the degree of Bachelor of Technology in Information Technology. It is also certified that the project has not been submitted or produced for the award of any other degree.

Dr. Shabir Ahmad Sofi
Supervisor
Dept. of Information Technology
NIT, Srinagar

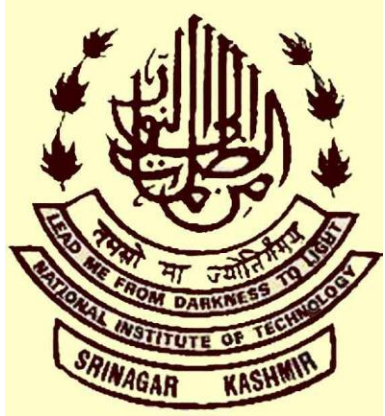

## STUDENTS DECLARATION

We, hereby declare that the work, which is being presented in the project entitled Analysis and Design of VANET Protocols for Srinagar City in partial fulfilment of the requirements for the award of the degree of Bachelor of Technology in Information Technology in the session 2019, is an authentic record of our own work carried out under the supervision of Dr. Shabir Ahmad Sofi, Department of Information Technology, National Institute of Technology, Srinagar. The matter embodied in this project has not been submitted by us for the award of any other degree.

Dated:13-06-2019

Name : Furqan Yaqub Khan

Signature: furkaan

# ACKNOWLEDGEMENT

I would like to express my special thanks of gratitude to my guide Dr. Shabir Ahmad Sofi as well as our Head of Department (Ms. Arooj Nissar who gave me the golden opportunity to do this wonderful project on the topic Analysis and Design of VANET Protocols for Srinagar City, which also helped me in doing a lot of Research and i came to know about so many new things, I am really thankful to them. Secondly I would also like to thank my parents and friends who helped me a lot in finalizing this project within the limited time frame.

<div align="right">

Furqan Yaqub Khan
IT/03/15

8th Semester

B.Tech, Information Technology

NIT Srinagar

</div>

# Abstract

Vehicular ad-hoc network (VANET) is subclass of mobile ad-hoc network which is vehicle to vehicle and vehicle to infrastructure communication environment, where nodes involve themselves as servers and/or clients to exchange and share information. VANET have some unique characteristics like high dynamic topology, frequent disconnections, restricted topology etc, so it need special class of routing protocol. To simulate the VANET scenarios we require two types of simulators, traffic simulator for generating traffic and network simulator. In this project I created a sample scenario of VANET for AODV, DSDV, DSR routing protocols. I have used SUMO for generating traffic mobility files and NS-3 for testing performance of routing protocols on the mobility files created using Traffic simulator SUMO.

# Contents

# LIST OF FIGURES

IV

# LIST OF TABLES

v

# CHAPTER 1

# INTRODUCTION

Today, Vehicular Ad-hoc Network (VANET) is one of the most emerging areas for the improvement of Intelligent Transportation System (ITS). VANET is a special form of MANET, where Mobile Ad-hoc Network (MANETs) are self-configuring network of mobile nodes connected by wireless links, while, VANET are distributed and self-assembling communication networks. A technology that uses moving vehicles as nodes to create a mobile network is termed as VANET. Here, node movement is restricted by factors like road course, encompassing traffic and traffic regulations. The primary goal of VANET is to provide road safety and other value added services such as email, audio/video sharing etc.

## 1. VANET

From the last decade, mobile communication techniques have transformed the automotive industry by providing anytime anywhere communication between different devices. This ease of communication allows exchange of valuable information between devices just on the go. The seamless exchange of information on real time bases has turned out to become a new paradigm in the industry. Correspondingly, the advances in the information technology and communication have easily supported the idea of communication between mobile devices [5]. Among these advancements, the concept of Vehicular Ad-hoc NETworks (VANET) came into limelight which has opened new possibilities to avail the use of safety applications. VANET refers to a network created in an ad-hoc manner where different moving vehicles and other connecting devices come in contact over a wireless medium and exchange useful information to one another. A small network is created at the same moment with the vehicles and other devices behaving as nodes in the network. Whatever information the nodes possess is transferred to all other nodes. Similarly all the nodes after transferring their set of data receive the data being transmitted by other nodes. After accumulating all of such data, nodes then work to generate useful information out of the data and then again transmit the information to other devices [2][4]. The communication between devices expands in such as way where nodes are free to join and leave the network i.e. it is an open network. The new vehicles being launched in the market are now coming with equipped on board sensors which make it easy for the vehicle to easily join and merge in the network and leverage the benefits of VANET.

VANET is a variation of MANET (Mobile Ad-hoc NETwork). MANET comprises of nodes which communicate without central network and where nodes are equipped with networking capabilities. VANET on the other side has emerged as a challenging and more liable class or variation of MANET. The freedom of nodes to enter or leave the network in VANET calls for different routing protocols than MANET .

This inter vehicle communication leads to passing and receiving of information so as to increase traffic efficiency, detect road conditions, decrease collisions, detect emergency situations and overall increase the efficiency of the network. VANET transfers the information to distant devices as well with the help of multi hops [6].

VANET can be characterized by following factors:

1. Dynamic topology- The speed and direction of vehicles changes constantly thereby resulting in high dynamic topology

2. Intermittent connectivity- Connectivity between devices changes very frequently like connection between two devices exchanging information can disconnect anytime. The reason behind frequent disconnection is high dynamic topology.

3. Mobility Patters: A large section of vehicles follow a certain patterns to move which is generally a function of traffic signals, speed limits, highways, streets, road conditions etc. These patterns when observed help in the creation of routing protocols for VANET.

4. Unlimited power and storage: It is assumed that the nodes in VANET are capable of possessing an unlimited amount of power as well as storage capacity. Therefore the nodes are free to exchange the data without the foundations of power consumption or storage wastage.

On board sensors: VANET assumes that the nodes are seldom equipped with on board sensors which are capable of transmission of information to other devices or nodes.

## 2. NS3

NS3 a tool for simulating the real world network on one computer by writing scripts in C++ or Python. Normally if we want to perform experiments, to see how our network works using various parameters. We don't have required number of computers and routers for making different topologies. Even if we have these resources it is very expensive to build such a network for experiment purposes.

So to overcome these drawbacks we used NS3, which is a discrete event network simulator for Internet. NS3 helps to create various virtual nodes (i.e., computers in real life) and with the help of various Helper classes it allows us to install devices, internet stacks, application, etc to our nodes.

Using NS3 we can create Point To Point, Wireless, CSMA, etc connections between nodes. Point To Point connection is same as a LAN connected between two computers. Wireless connection is same as WiFi connection between various computers and routers. CSMA connection is same as bus topology between computers. After building connections we try to install NIC to every node to enable network connectivity.

When network cards are enabled in the devices, we add different parameters in the channels (i.e., real world path used to send data) which are data-rate, packet size, etc. Now we use Application to generate traffic and send the packets using these applications.

## 3. PROTOCOLS

Wireless network can be classified into infrastructure based and infrastructure less network. In the case of infrastructure based networks, Access point are used for communication. They act as routers for the nodes within their communication range. Whereas, in infrastructure less networks, also known as, ad hoc networks, nodes act as routers. A Mobile ad hoc network (MANET) is a type of ad hoc network in which nodes can change locations.

The routing protocols in MANET are broadly classified into three categories, namely, proactive protocols, reactive protocols, hybrid protocols. Proactive protocols, also known as table-driven protocols, maintain routing information in the routing table of each node. The proactive routing protocols are Destination-Sequenced Distance-Vector (DSDV) routing protocol. The reactive protocols are Ad-hoc On-demand Distance Vector (AODV), Dynamic Source Routing (DSR).

**A. Ad-hoc On-demand Distance Vector (AODV)**

AODV is a combination of on-demand and distance vector i.e. hop-to-hop routing methodology. When a node needs to know a route to a specific destination it creates a ROUTE REQUEST. Next the route request is forwarded by intermediate nodes which also create a reverse route for itself for destination. When the request reaches a node with route to destination it creates again a REPLY which contains the number of hops that are require to reach the destination. All nodes that participate in forwarding this reply to the source node create a forward route to destination. This route created from each node from source to destination is a hop-by-hop state and not the entire route as in source routing.

**B. Destination Sequenced Distance Vector (DSDV):** DSDV is a hop-by-hop distance vector routing protocol requiring each node to periodically broadcast routing updates based on the idea of classical Bellman-Ford Routing algorithm [2]. Each node maintains a routing table listing the "next hop" for each reachable destination, number of hops to reach destination and the sequence number assigned by destination node.

The sequence number is used to distinguish stale routes from new ones and thus avoid loop formation. The stations periodically transmit their routing tables to their immediate neighbors. A station also transmits its routing table if a significant change has occurred in its table from the last update sent. So, the update is both time-driven and event-driven. The routing table updates can be sent in two ways: a "full dump" or an "incremental".

**C. Dynamic Source Routing(DSR):** DSR is a simple and efficient protocol designed specifically for use in multiple wireless adhoc networks of mobile nodes. It allows nodes to dynamically discover a source route across multiple network hops to any destination in adhoc network. Each data packet sent then carries in its header the complete ordered list of modes through which packet must pass, allowing packet routing to be a trivially loop free and avoiding the need for up-to-date routing information in the intermediate nodes through which the packet is forwarded. With the inclusion of this source route in the hearder of each data packet, other nodes forwarding or overhearing any of the packets may easily cache this route information for future use.

# CHAPTER 2

# LITERATURE SURVEY

1. **SUMO**

**1.1 Contributors and Participants**

| Org. | Name | Topics / Contribution |
|---|---|---|
| 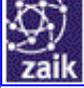 | Christian Rössel | Initial microsimulation core; initial detectors implementation |
| 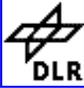 | Peter Wagner | Models, organisation, spiritual lead |
| | Daniel Krajzewicz | Everything |
| | Julia Ringel | Traffic Light & WAUT Algorithms |
| | Eric Nicolay | Everything |
| | Michael Behrisch | Everything |
| | Yun-Pang Wang | User Assignment |
| | Danilot Teta Boyom | Vehicular Communication Model (removed from the source) |
| | Sascha Krieg | |
| | Lena Kalleske | |
| | Laura Bieker | Tests, Python scripts |
| | Jakob Erdmann | network import, NETEDIT. |
| | Andreas Gaubatz | |
| | Maik Drozdzynski | |
| Uni Lübeck | Axel Wegener | TraCI initiator |
| | Thimor Bohn | TraCI |
| | Friedemann Wesner | TraCI |
| | Felix Brack | |
| | Tino Morenz | |
| 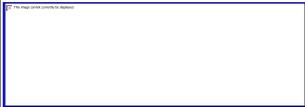 | Christoph Sommer | TraCI merge with Veins, Subscription Interface, Misc. |
| 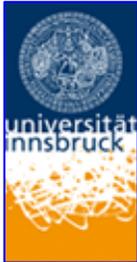 | David Eckhoff | TraCI, deterministic simulation behavior |
| | Falko Dressler | TraCI |
| 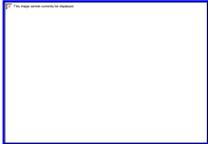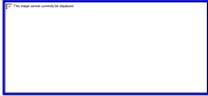 | Tobias Mayer | Traffic model abstraction, IDM model port |
| HU Berlin | Matthias Heppner | Unittests |
| 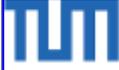 | Piotr Woznica | ACTIVITYGEN |
| | Walter Bamberger | Development of ACTIVITYGEN  as a base for the evaluation of trust scenarios in VANETs. The work is part of the project Fidens: Trust between cooperative systems featuring trusted probabilistic knowledge processing in vehicular networks. |
| | | |
| IIT Bombay, India | Ashutosh Bajpai | randomDepart.py, a python script to generate the real traffic pattern by exponential Distribution. |
| 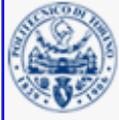 | Enrico Gueli | TraCI4J |
| | Leontios Papaleontiou | Sumo Traffic Modeller |
| 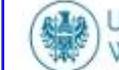 | Karol Stosiek | Documentation, network building |

Table - 1

2. **NS3 and VANET Protocols.**

**Chia-Chen Hung et al (2008)**, had analyzed and demonstrated ,the Intelligent Transportation System (ITS), a worldwide initiative program that utilized novel information and communication technology for transport infrastructure and vehicles. Among extensive ITS components, efficient communication system was the most important role that connects numerous vehicles with roadside infrastructure and management center in the ITS program. In this paper a new Heterogeneous Vehicular Network (HVN) architecture and a mobility pattern aware routing(MPARP) for HVN was proposed. HVN integrates Wireless Metropolitan Area Network (WMAN) with VANET technology and reserves advantages of better coverage in WMAN and high data rate in VANET. Vehicles in HVN can communicate with each other and access Internet ubiquitously.

**T. Nisha Devi and Adiline T. Macriga (2010)**, had presented about today's communication industry that concentrates more on the live information transfer without altering the existing infrastructure and hence required a single convergence platform of all networks' access. The next generation systems support multimode, multi-access and reconfigurable devices to support inter-working of heterogeneous networks. The network selection is user-centric and based on multiple QOS (Quality of Service) parameters like bandwidth, cost, security level, call drop probability etc., to select appropriate networks. The proposed algorithm used a distance function to generate an ordered list of various access technologies called networks in a particular region according to multiple user preferences and level of interest. Further level of customization was done with user preference in terms of giving priority to few parameters and was implemented by weighted distance function.

**Benslimane Abderrahim et al (2011)**, had presented coupling the high data rates of IEEE 802.11p-based VANETs and the wide coverage area of 3GPP networks (e.g., UMTS), that envisions a VANET-UMTS integrated network architecture. In this architecture, vehicles are dynamically clustered according to different related metrics.

From these clusters, a minimum number of vehicles, equipped with IEEE 802.11p and UTRAN interfaces, are selected as vehicular gateways to link VANET to UMTS. Issues pertaining to gateway selection, gateway advertisement and discovery, service migration between gateways (i.e., when serving gateways lose their optimality) are all addressed and an adaptive mobile gateway management mechanism was proposed. Simulations were carried out using NS2 to evaluate the performance of the envisioned architecture incorporating the proposed mechanisms.

**Mohamad Yusof Darus and Abu Bakar Kamalrulnizam (2011)** had studied some of the Congestion control algorithms in Vehicular Networks (VANETs) The study further exposed the weaknesses and advantages of some of these congestion control algorithms which could assist researchers to tackle the inherent problems of congestions in VANETs. This paper also concluded with a planned future research for disseminating uni-priority of event-driven safety messages while solving congestion problems.

**Kalyani B. Amit (2012),** had analyzed, a heterogeneous network framework for providing seamless connectivity and data transfer between various network technologies. This framework provides QoS seamless heterogeneous network architecture. The heterogeneous network of wireless communication was expected to integrate potentially a large number of heterogeneous wireless technologies which could be considered a huge step forward towards a universal seamless access. One of the main challenges for seamless mobility was the availability of reliable horizontal (intra system) and vertical (inter system) handoff schemes.

**Singhal Manav and Shukla Anupam (2012)**, had proposed the implementation of Location based services through Google Web Services and Walk Score Transit APIs on Android Phones to give multiple services to the user based on their Location. Location based Services offer many advantages to the mobile users to retrieve the information about their current location and process that data to get more useful information near to their location. With the help of A-GPS in phones and through Web Services using GPRS,

Location based Services can be implemented on Android based smart phones to provide these value-added services: advising clients of current traffic conditions, providing routing information, helping them find nearby hotels.

**Sharma Manish and Gurpadam Singh (2012)**, had discussed various ad-hoc routing protocols, Reactive, Proactive & Hybrid, taking in to consideration parameters like speed, altitude, mobility etc in real VANET scenario. The AODV and DYMO (Reactive), OLSR (Proactive) and ZRP (hybrid) protocols were compared for IEEE 802.11(MAC) and IEEE 802.11(DCF) standard using Qualnet as a Simulation tool, Since IEEE 802.11, covered both physical and data link layer. Hence performance of the protocols in these layers helped, to make a right selection of Protocol for high speed mobility. Varying parameters of VANET showed that in the real traffic scenarios proactive protocol performs more efficiently for IEEE 802.11 (MAC) and IEEE 802.11(DCF).

**Muhammad Rizwan Arshad et al (2012)**, had examined, WiMAX and WiFi on Vehicular Ad-hoc Networks (VANET) which were used to evaluate the best service provider technology for VANET. In VANET the nodes are moving very fast and change their network infrastructure rapidly, which have very short time to communicate with each other. Both WiMAX and WiFi is to be used as per their features in the long distances areas and then their practice in real model. The focus of this research was to reduce the delay time of message passing, authentication and to find the best suitable and qualitative service from WiMAX and WiFi.

**Gadkari .Y Mushtak and Sambre .B Nitin (2012)**, had made an attempt for identifying major issues and challenges associated with different VANET protocols, security and simulation tools. Security and privacy are indispensable in vehicular communications for successful acceptance and deployment of such a technology. The vehicular safety application should be thoroughly tested before it is deployed in a real world to use. Simulator tool has been preferred over outdoor experiment because it simple, easy and cheap. The increasing popularity and attention in VANETs has prompted researchers to develop accurate and realistic simulation tools. In this paper, we make a survey of several publicly available mobility generators, network simulators, and VANET simulators was made.

**Patel Nikhil and Parmar Kiran (2012)**, had presented several optimizations methods for the execution of vertical handoff decision algorithms, with the goal of maximizing the quality of service experienced by each user and a method to select the handover target network . Future wireless networks must be able to coordinate services within a diverse-network environment. One of the challenging problems for coordination is vertical handoff, which is the decision for a mobile node to handoff between different types of networks. While traditional handoff is based on received signal strength comparisons, vertical handoff must evaluate additional factors, such as monetary cost, security, power consumption, network conditions, and user preferences. In this paper, the method proposed is a combination of weight distribution and cost factor calculation. Weights of various network parameters are generated based on user preferences and the power level of mobile terminal, and cost factors of candidate networks are calculated using a cost function. The network with the Highest $Q_i$ and lowest cost is selected as the handover target network. This method was able to maximize the user's satisfaction level by choosing the one access network.

**P.Vetrivelan et al (2012)**, had suggested a Multi-Constraint Realtime Vehicular (MCRV) mobility framework that is equipped with important criterion like collision avoidance between vehicles and traffic reports. Also certain vehicles such as ambulance, fire service vans, police patrols needed to be given a high priority in our envisioned network architecture, as their requirements are crucial during emergency situations. Hence, enabling QoS for differentiating the services according to vehicular priorities and providing group communications, alongside vehicular collision avoidance, was implemented using NS3 and SUMO. The vehicle-to-vehicle (V2V) and vehicle-to-infrastructure (V2I) communications were done using WAVE and WiMAX/UMTS heterogeneous networks respectively. The horizontal and vertical handovers were chosen at appropriate rite decision for their communications.

**Bhat .S Vijaylaxmi and Shah .N Pragnesh (2012)**, had studied, the problem of multihop routing in vehicular ad hoc networks (VANET). IEEE 802.11p and other vehicular network standards advocate vehicles to issue periodic broadcast messages at regular intervals called beacons. The physical rate adaptation in 802.11 was deeply investigated, though still open issue. Since 802.11 used the random access Distributed Coordination Function (DCF) mechanism to access the medium, collisions could occur when two or more stations wanted to transmit data simultaneously. In this paper the authors had proposed a rate adaptation algorithm that behaved like Auto Rate Fallback (ARF), but made use of the RTS/CTS handshake, when necessary, to decide whether the physical transmission rate should be changed. The performance of rate adaptation algorithm, was then compared with other well known algorithms, AODV and DSDV.

**Yuvraj Singh(2012)**, had suggested that, the radio propagation was essential for emerging technologies with appropriate design, deployment and management strategies for any wireless network. It is heavily site specific and can vary significantly depending on terrain, frequency of operation, velocity of mobile terminal, interface sources and other dynamic factor. Large scale path loss modeling plays a fundamental role in designing both fixed and mobile radio systems. Predicting the radio coverage area of a system is not done in a standard manner. Therefore, before setting up a system, one has to choose a proper method depending on the channel's BTS antenna height gain to show good result.

**Mor Annu (2013),** had discussed about, Vehicular Ad Hoc Network (VANET) which is a sub class of mobile ad hoc networks. VANET is most advanced technology for intelligent transportation system that provides wireless communication among vehicles and vehicle to road side equipments, according to IEEE 802.11p standard for end to end communication between vehicles . One of the most important routing protocols used in ad hoc networks was AODV. This protocol is connectivity based reactive protocol that searches routes only when they are needed because bandwidth is limited and topology frequently changed. It always exchanges control packets between neighbor nodes for routing. In this article author presented cross layer technique that found channel security at link layer to AODV routing protocol to improve the communication in vehicles for safety purpose. To reduce the packet delay in AODV , the routing protocol (AODV_BD),was proposed. It reduced the packet delay in AODV and made routes more stable.

**Burde Ashwini and S. P. Pingat (2013)** , had focussed on the use of VANET technology for efficient traffic management and route planning while vehicle heads from source to destination. VANET technology was used as a medium to generate updated information for the vehicle when it headed from source to its destination. Performance of SA, PSO and CA algorithm about traffic alerts were also compared.

**Sandhu K. Gurveen et al (2013)**, had tried to discuss two latest wireless technologies: Wi-Fi and WiMAX. Option way out to the trouble of accessing information in remote areas where wired network are inaccessible was offered by Wireless Networking Technology. Wireless Networking had changed the way people communicate and share information by eliminating the boundaries of distance and location.

**K. Komala and Dr. P. Indumathi (2013),** had discuss about, A Heterogeneous Network abbreviated as HetNet , which is a mix of macrocells, picocells,femtocells, remote radio heads and relays. A HetNet is a network consisting of various wireless access technologies and functionalities. The Next Generation mobility technique called the 4G deals with the multi-access heterogeneous wireless networks which provide seamless connectivity of multimedia services at a higher data rate to the end users. There are various research challenges like self organization, backhauling, handover and interference for the 4G. The various handover mechanisms with respect to intra-domain and inter-domain are analyzed. An experimental testbed for the seamless mobility of heterogeneous wireless techniques such as mobility from IEEE 802.11 to IEEE 802.16e was considered. The performance results of a seamless handoff with minimum packet loss and delay proved the efficiency in the mobility of the HetNets using the Network Simulator (NS-2).

**Ghonge M. Mangesh and Gupta G. Suraj (2013)**, had described WLAN, WPAN and WiMAX technologies that were introduced and comparatively studied in terms of peak data rate, bandwidth, multiple access techniques, mobility, coverage, standardization, and market penetration. Detailed technical comparative analysis between WLAN, WPAN, WiMAX wireless networks that provide an alternative solution to the problem of information access in remote inaccessible areas where wired networks are not cost effective had been looked into. Their work had proved that the WiMAX standard goal was not to replace Wi-Fi in its applications, but rather to supplement it in order to form a wireless network web.

**Md. Humayun Kabir (2013)** had presented the aspects related to Vehicular Ad Hoc Network (VANET) , which is a kind of special wireless ad hoc network, that has the characteristics of high node mobility and fast topology changes. VANET has become an active area of research, standardization, and development because it has tremendous potential to improve vehicle and road safety, traffic efficiency and convenience as well as comfort to both drivers and passengers. Vehicular networks will not only provide safety and lifesaving applications, but they will become a powerful communication tool for their users and help researchers and developers to understand and distinguish the main features surrounding VANET.

**Jain Sapan Kumar and Badhe Vivek (2013)**, had presented about, how a Wireless Sensor Network varies from Heterogeneous Network. In the wireless sensor network , hundreds of sensor node are used. If energy, range and hardware capabilities are different in various nodes in the network then , these types of network are mainly known as heterogeneous network. Energy in the sensors is a scarce resource . It must be managed in an efficient manner to expand the life of network and a secure multi-hop reactive protocol for heterogeneous wireless sensor network with clustering was designed.

**Gaur Saurabh Kumar et al (2013)**, had analyzed the critical factors in deciding the networking framework over which the future vehicular applications would be deployed. A reactive research effort is needed for making VANETs a reality in the near future. A vehicular Ad hoc network (VANETs) can be used as an alert system. By this we get the alert about the traffic jam. It helps to create balance in traffic load to reduce travelling time. This system is also useful to broadcast emergency signal to the driver of the vehicle behind the accident. It also helps to send message to ambulance and traffic police in the case of traffic emergency.

**Jaiswal Siddhant and Dr D. S. Adane (2013)**, had provided a routing algorithm which works on a hybrid scenario, i.e. it will have both static and dynamic infrastructure. The approach used was Cluster based routing which will help in transmitting packets even in a network with low vehicle density.

**Tejpreet Singh et al (2013)**, had discussed about VANETs ,that are highly dynamic in nature due to mobility of nodes and this dynamic nature caused topological change in the network, which may affect the communication and security of whole network.There are various attacks which may effect the network, but wormhole attack is one the harmful attack which may affect the communication in VANET. This is so because wormhole may lead to attacks like Denial of service attack, data tampering, masquerading etc. In this paper performance of different routing protocols were analysed on the basis of metrics like throughput, end-to-end delay and jitter. Performance of routing protocols were analysed in two cases first , without wormhole attack and second is with wormhole attack and it has been checked how much performance of routing protocols AODV, OLSR and ZRP was degraded with wormhole attack.

**Vishal Kumar et al (2013)**, had primarily categorized the various possible applications of vehicular network, along with its features, and implementations in the real world. The applications of VANETs are of the classes :1) Safety oriented, 2) Commercial oriented 3) Convenience oriented and 4) Productive Applications.

**E Abinaya and R Sekar (2014)**, had proposed an idea to optimize signal control at traffic intersections which used vehicular ad hoc networks (VANETs) to collect and aggregate real-time speed and position information on individual vehicles. An online algorithm, referred to as the oldest job first (OJF) algorithm was used to minimize the delay across the intersection. The results were compared with vehicle-actuated methods ,Webster's method ,and pre-timed signal control methods .

**Mathew Ann Bittu, Joseph Sumy (2014)**, had discussed about ,Heterogeneous network, which is an important component of cellular networks to meet the increasing mobile data demands. Due to uneven traffic distribution, some nodes suffer from heavy load, and their adjacent nodes may carry only light load. This load imbalance among nodes restricts the network from fully utilizing its capacity and providing better services to users. So there is a need for load balancing mechanism to be present in the network. In this paper a load balancing scheme was proposed that moves the load of heavy nodes to lightly loaded or idle nodes by finding next shortest path. This may lead to efficient utilization of nodes and nodes can allocate resource efficiently.

**Jaswal Kamini et al (2014)** , had studied the selection of a network simulator for evaluating research work. According to the previous researches on the performance of OPNET, it had been proved that OPNET is comparatively more reliable, easy to understand and implement network simulation tool than its other counterparts. The main focus of this paper was to study the Wimax performance when implemented with OPNET. OPNET is based on a mechanism called discrete event system which means that the system behaviour can simulate by modeling the events in the system in the order of scenarios the user has set up. Hierarchical structure is used to organize the networks.

**Anwer M. Shahid and Guy Chris (2014)**, had surveyed some of the key vehicular wireless access technology standards such as 802.11p, P1609 protocols, Cellular System, CALM, MBWA, WiMAX, Microwave, Bluetooth and ZigBee which served as a base for supporting both Safety and Non Safety applications. The survey also analysed and compared the wireless standards using various parameters such as bandwidth, ease of use, upfront cost, maintenance, accessibility, signal coverage, signal interference and security. Finally, the work discussed some of the issues associated with the interoperability among those protocols.

**Mohammed Shafeeq Ahmed (2014)**, had discussed and illustrated, security solution, various vulnerabilities and possible attacks to WiMAX network. In IEEE 802.16, security has been considered as the main issue during the design of the protocol. However, security mechanism of the IEEE 802.16 (WiMAX) still remains a question. WiMAX is relatively a new technology; not deployed widely to justify the evidence of threats, risk and vulnerability in real situations. This paper addressed, the security aspects of the IEEE 802.16 Standard and pointed out the security vulnerabilities, threats and risks associated with both layers in WiMAX physical and MAC Layers. The threats apply to both layers of WiMAX. At PHY layers, jamming can be considered a major threat. At MAC layer, critical threats include eavesdropping of management messages, masquerading, management message modification or DoS attacks.

**Pooja Rani et al (2014)**, had surveyed different outdoor and indoor propagation models. In wireless communication, path loss was caused by different obstacles between the transmitter and receiver that absorb power due to which signal strength is reduced. To calculate the path loss between the transmitter and receiver, different path loss models are used like okumara, hata, cost 231 etc. These path loss models may give different results in urban, suburban and rural areas. These models depend on various parameters like mobile-station antenna height, transmitter-receiver distance, base - station antenna height.

**T.Karthikeyan and B. Subramani (2014),** had surveyed about QoS based agent routing algorithms in MANET, WSN and VANET. One of the most challenging tasks in Ad-hoc Network(MANET &VANET) is Quality of Service (QoS) which is determined by numerous parameters such as bandwidth and delay constraints, varying channel conditions, power limitations, node mobility, dynamic topology, packet delivery ratio, end-to-end delay and connection duration. With the increasing demand for real time applications in the Wireless Senor Network (WSN), real time critical events anticipate an efficient quality-of-service (QoS) based routing for data delivery from the network infrastructure. Designing such QoS based agent routing protocol to meet the reliability and delay guarantee of critical events while preserving the energy efficiency was a challenging task.

**H .Vishalakshi Prabhu, G.S.Nagaraja (2014),** had surveyed on the Worldwide mobile operators, industry experts, and researchers that have diverse visions of potential 4th generation (4G) features and its implementations. This paper had given a survey and classification of the important QoS approaches proposed for 4G networks. Classification was based on the work done in each protocol layer and Cross Layer Design (CLD) approach. Finally, this paper presented outcomes of survey which included significant observations, limitations and idea of further research in improving QoS in 4G networks.

**Anuradha Singh and Mintu Singh (2015),** have discussed the overview of Vehicular Ad-hoc Network (VANET), which is a most critical class of mobile ad-hoc network (MANET) that enables intelligent communication among vehicles and also between vehicle and roadside infrastructures. It is a promising approach for the Intelligent Transport System (ITS). There are many challenges to be addressed when employing VANET. It has a very high dynamic topology and constrained mobility which makes the traditional MANET protocols unsuitable for VANET. The aim of this review paper was to give an overview of the vehicular ad hoc networks, its standards, applications, security issues and the existing VANET routing protocols.

**Kshirsagar Suresh Nikhil and Dr. U. S. Sutar (2015)**, have thrown light on accident prevention and traffic signal control for ambulance, police van, and normal vehicles too. To overcome this they have implemented a highway model, intersection model that manages vehicle mobility and shows the actual communication between vehicle to vehicle (V2V) and vehicle to infrastructure (V2I).The security of VANET technology is one of the most critical issues because their information transmission is propagated in open access environments. Over a period of years, VANET has received increased attention as the potential technology to enhance active and preventative safety on the road.

**Sadek M. Noha et al (2015),** have discussed about Intelligent Transportation Systems (ITS) that have been receiving significant interest from various stakeholders worldwide. ITS promise major enhancements to the efficiency, safety, convenience and sustainability of transportation systems. To satisfy the diverse vehicular application requirements, this paper had proposed, an integration of IEEE 802.11-based VANET and LTE cellular network using mobile vehicular gateways. IEEE 802.11 g is used for V2V communications and LTE for V2I communications. A burst communication technique is applied to prevent packet losses in the critical uplink ITS traffic. A performance simulation-based study was conducted to validate the feasibility of the proposed system in an urban vehicular environment. The system performance was evaluated in terms of data loss, data rate, delay and jitter. The results indicated that the proposed Multi-RAT system offers acceptable performance that meets the requirements of the different vehicular applications.

**Chib Randeep Singh et al (2015)**, have suggested that Radio wave propagation models are extremely important in radio network planning, design as well as in interference planning. Radio propagation is essential for emerging technologies with appropriate design, deployment and management strategies for any wireless network. It is heavily site specific and can vary significantly depending on terrain, frequency of operation, velocity of mobile terminal, interface sources and other dynamic factor. Accurate characterization of radio channel through key parameters and a mathematical model is important for predicting signal coverage, achievable data rates, BER and Antenna gain. Path loss models for macro cells such as Okumura, Hata and COST 231 models were analyzed and compared with their parameters.

# CHAPTER 3

# THEORY

A routing protocol defines the way the dissemination of a message in a network is handled. It defines the creation of a route from the source of the message its destination. In Ad-Hoc networks, every node is responsible to deal with the routing and do not rely on specified devices like network with infrastructure. This part will firstly identify several ways to classify routing algorithms, secondly, it will present the routing protocols available in VANETs. Finally, a table will summarize this chapter by classifying the algorithms according to their different routing techniques. In the next section, I would introduce important routing algorithms.

## 2.1 Unicast, Multicast and Broadcast

Those three types of communication define the number of point concerned by a transmitted message in a network. Unicast describes a communication between two points, the sender and the receiver. Multicast defines a communication between a sender and several receivers. Finally, broadcast describes a communication where the sender's message is sent to all the other nodes of the network. In VANETs, the three types of communication are used, depending on the types of application. Environment and entertainment application will use more often unicast and multicast because the message does not concern every vehicle. On the other Hand, safety application will mainly use broadcast communication to reach all nodes of the network. Unicast and multicast transmissions need the establishment of a communication between the sender and the receiver(s) before the transmission. The absence of routing devices (such as gateway) in ad-hoc network implies that the communication will consist of a succession of hops from the sender to the receiver(s). Where each node in the path will forward the message until it reaches its destination(s). The way the algorithm defines this path is called route discovery and is the first phase for most of unicast and multicast protocol in VANETs. Most of the time it uses broadcasting methods during this phase. Therefore, even if this thesis focusses more on safety application and on broadcasting algorithms, we will also go through some unicast and multicast protocols that use broadcasting techniques in their first phase. Finally, the results obtained for a new broadcasting algorithm could also impact the efficiency of unicast and multicast algorithm by improving the route discovery phase.

## 2.2 Proactive, Reactive and Hybrid protocols

Routing protocols can be classified under different criteria. From the way it handles routes in the network to the way it discovers those routes. In this section, we identify three behaviors defining when a route is established and maintained in the network. Those three types are proactive protocols, reactive protocols and hybrid protocols and will be detailed below.

### 2.2.1 Proactive Protocols

Proactive protocols use a route discovery phase before sending any data. Indeed, routes are calculated and maintained up to date continuously by transmitting periodic routing information on the network. Each time a node wants to transmit a packet, the packet is sent only if a route to the destination is established. Otherwise the packet will wait in queue until a route has been found. Those types of protocols are difficult to maintain in highly dynamic and scalable network such as VANETs. Therefore, they require a significant amount of routing information to be transmitted, increasing significantly the bandwidth consumption. Accordingly, Proactive protocols are not the most suitable for VANETs. Examples of proactive protocols such as Destination Sequenced Demand Vector (DSDV) is detailed later in this chapter.

### 2.2.2 Reactive Protocols

For reactive protocols, the route discovery phase is initiated only when a packet needs to be sent over network. If a route is found it will be maintained using maintenance route packets sent periodically until the destination is not reachable. Reactive protocols present lower overhead than proactive protocols, but the end to delay is more important due to the route discovery phase started every time a packet needs to be send. Therefore, reactive protocols present also some inconvenient in VANETs where the transmission require a low end to end delay. Examples of reactive protocols such as Ad-hoc On Demand Distance Vector (AODV), Distance Vector Routing (DSR) are detailed later in this chapter.

### 2.2.3 Hybrid Protocols

Hybrid protocols are designed to compensate the overhead of proactive protocols and the end to end delay of reactive protocols by combining properties of both types of protocols. In most of hybrid protocols in VANETs, each node of the network broadcast its routing information (using beacon messages). Thus, nodes create and maintain a table of neighbors based on those beacons. The route discovery is then initialized when a packet needs to be sent and use the neighbors' tables to find destination faster. Hybrid protocols have been made to handle dynamic networks such as Mobile Ad-Hoc Networks (MANETs) and then modified to fit the high speed and scalability of VANETs. Examples of Hybrid protocols such as Zone Routing Protocol (ZRP) and Temporarily Ordered Routing Algorithm (TORA) are given in more details later in this chapter.

### 2.3 Categorization of next hop selection

In routings protocols in MANET and VANET, the route discovery phase uses different methods to select the next node in the route (called next hop). Indeed the selection of the next hop is made on several criteria depending on the method used. We distingue height types of next hop selection which are further node, best quality link, most demanding node, probability base, backbone node, stochastic method, counter based, distance to mean. Those method are not necessary independent and are explained below.

### 2.3.1 Distance-based

The distance-base technique consists of selecting the next hop based on the distance between the current node and the closest to the destination. This technique can either rely on node geographical location especially in VANET but can also rely on the network topology. An important amount of algorithm uses this technique during the route discovery phase named greedy forwarding. In greedy forwarding, each node selection the next hop the closest geographically to the destination (This technique is called furthest node selection). In broadcasting routing protocols where the destination position is not known, each node selects the node the furthest away from itself in its transmitting range considering that close nodes will already have received the message. The furthest node can also consist of a of each node decide by itself whether to forward the message by comparing its distance to the precedent hop and comparing it to a threshold value. Examples of Routings protocols using greedy forwarding and further nodes are contained in geographical routing detailed in this chapter.

### 2.3.2 Best Quality Link

Selecting the next hop based on the quality of the link means considering real word channel conditions. This selection may rely on distance with next hop, the received power from beacon packets or other channel criteria. This type of protocols can increase the end-to-end delay because of the information gathering phase. However, it shows a better reliability than furthest node selection that ignores channel condition.

### 2.3.3 Most Demanding node

The most demanding node method prioritizes certain nodes of the networks according to their locations in the graph or their time to react to previous messages. The idea is to transmit the message to the nodes the most concerned by it. This type of method is mainly used in warning message propagation in VANET to provide security relative information to the network. These type of selection ensure the delivery of important packets to demanding nodes. However, it increases calculation considerably to identify the demanding area and therefore the end to end delay.

### 2.3.4 Probability based forwarding

Each node will forward the packet depending on a certain probability. This is used to decrease the number of packet rebroadcasting a packet. The assigned probability is either defined or dynamically change depending on the network density or node location. This method is mainly used for some broadcasting protocols where all nodes of the network are concerned by the message so the objective is to decrease the number of rebroadcasting nodes.

### 2.3.5 Backbone node

Backbone node selection consider the existence of infrastructure in the network. Those can be physical, using road side units for examples like in some trajectory based routing protocols. Certain can also be dynamically created by forming groups of nodes where some of them will be entitled to the routing (like in cluster based routing algorithm). Those infrastructures will be used to either calculate the route to destination or to select next forwarding node. The goals of this selection is to decrease the overhead by allowing only backbone nodes to transmit However, the node selection requires calculation that are needed to be kept as simple as possible to maintain a reliable end to end delay.

### 2.3.6 Stochastic method

In stochastic method, the next hop is selected randomly among the neighbors available in transmitter based algorithm. In receiver based algorithm, each node selects randomly either to forward or not the packet. This type of selection do not ensure the packet delivery and is therefore not reliable in route request or broadcast routing.

### 2.3.7 Counter based

The counter based method consists on counting the number of time a message is received by a node. When a node receives a message, it will set up a time to wait before forwarding the message and if it receives again the message before this time is over, it will cancel the forwarding. This defined time is called a back off timer and can be calculated over some parameters such as node location for example. This solution can be most of the time used with another selection technique sur as distance-based for example in the form of a back off timer. The Counter-based method is mainly used in geographic broadcast routing protocol to propagate message to all node and decreasing overhead.

However, it can also be used to create route to destination among the first node to forward to the destination.

### 2.3.8 Distance to mean

The Distance to mean selection can be compared to the distance-based technique and is mostly a receiver-based method, where each node decides by itself to forward the message. However, here the decision is made on the distance between the node and the spatial mean of precedent forwarders of the packet. This value is then compared to a threshold value calculated over several parameters. This selection method shows better results than distance based ones such as further node in terms of reachability, bandwidth consumption and end to end delay [1]. Therefore, the algorithm proposed on this paper will use this type of next hop selection. Algorithms based on this technic will be detailed in more depth in the next sections of this chapter.

### 2.4 Types of routing protocols

Now that three types of communication have been distinguee, the routing protocols based on those communication can be separated into six categories. We identify topology based, location based, cluster based, Geocast based, trajectory based and geographic based routing. All those categories are sorted by the way the transmitted message is conveyed to its destination.

### 2.4.1 Topology Based Routing

Topology based routing consist on the establishment and the maintenance of routing table for each node of the network. It means that every node in the network knows the path to reach other node in the network. This type of protocol is not fitted for VANET because of the high mobility and scalability of vehicular network. Maintaining such table is hard when the topology of a network is constantly changing. However, certain protocol adapts such protocols to make it more efficient in VANET environment.

### 2.4.1.1 AODV

Ad-hoc On-Demand Vector (AODV) routing is a neighborhood awareness protocol because each node uses Beacon hello messages to keep track of its neighbors. AODV is also a reactive routing protocol because the route generation is started only when a node wants to send a packet. Then, a Route Request (RREQ) is sent to neighbors and propagated between net-hop neighbors to find a path to destination. Then, a Route Reply (RRep) is sent back to the source using the reverse route. AODV control if those routes contain no loop and find the shortest path among them. Each node then stores next hop to destination in a table. It can also create new routes or modify existent by handling errors.

AODV is one of the most used protocol in wireless networks because of its viability, however, it is not fit to handle high mobility and scalability among VANET. Therefore, a lot of protocol used in VANET are adapted from AODV with modification to fit more VANET the needs. Finally, AODV provide Unicast and Multicast by establishing routes to destination which is node needed when you send warning messages in VANET which only needs Broadcasting.

### 2.4.1.2  MAV-AODV

Multicast with Ant Colony Optimization of AODV (MAV-AODV)  is based on AODV and is a Bio inspired algorithm. Based on Ant colony, MAV-AODV use nodes' mobility information to build multicast tree and sustain its lifetime. A neighbor table is maintained using periodic beacons to obtain mobility information. Then the Request Response phase of AODV is improved using node's position and a Best quality link next hop selection.

### 2.4.1.2 OLSR

Optimized Link State Routing (OLSR) Protocol is a proactive protocol because it creates and maintain a routing table based on topology information regularly exchanged by the nodes. Then certain nodes are classified as Multi-Point Relay (MPR) by their neighbors. Information that they broadcast periodically. Those nodes are then used to form the route from a node to the destination. Like AODV, maintaining a routing table in VANET networks is not efficient because of the ephemeral state of the network but also in terms of bandwidth consumption. Then, OLSR also provide Unicast and Multicast which are not needed for warning message propagation in VANET.

### 2.4.1.3 DSDV

Destination Sequenced Distance Vector (DSDV) uses routing table scheme where the path is calculated based on the Bellman-Ford algorithm which is, in graph theory, an algorithm to find the shortest path form a single vertex to other vertices in a weighted digraph. The main used of this algorithm is to avoid the routing loop problem.

### 2.4.1.4 DSR

Like AODV, Dynamic Source Routing (DSR) starts route discovery operation when a node wants to send a message. But unlike AODV, the route is kept in full in the table and maintained for a period. Likewise, the RREQs are made using flooding, not by using a maintained table of neighbors. Furthermore, every node is responsible to wait for reception confirmation from next hop. Until then, it will keep sending the packet until it reaches a defined maximum threshold.

### 2.4.1.5 TORA

TORA is alike DSR but in addition to the route discovery phase, this protocol provides a route erase phase. Thanks to the first phase, every node constructs and maintains a route table. Nodes are then able to detect network partition, in that case, they will trigger the erase phase by sending a clean packet that will delete the invalid route.

### 2.4.1.6 FSR

FSR uses flooding broadcasting to propagate packets in the networks. Then, with the latest location information contains in those messages, each node builds and maintains a Topology Table which allows node to build routes.

### 2.4.1.7 ZRP

In ZRP, every node's neighborhood is delimited by a defined zone depending on the transmission range. For nodes inside this area, routes are discovered reactively. However, to transmit to nodes outside of the transmission zone, a route request is emitted to other zones. ZRP is a hybrid protocol, between topology based and cluster based because nodes are grouped in zone. But in Cluster based routing algorithm a cluster head is designed to deal with all routing inside and outside of the cluster.

### 2.4.1.8 MAR-DYMO

Mobility-aware Ant colony optimization Routing (MAR-DYMO) is an integration in Mobile Ad-Hoc Networks (MANET) of Ant Colony Optimization (ACO) [20] by combining it to the routing protocol Dynamic MANET On-demand (DYMO). DYMO is the successor of AODV protocol and is based on the same principle of multi-hop propagation between neighbors until it reaches destination. ACO works with several principles. One of them is the pheromones which are used to grade route to increase reliability. In MAR-DYMO, more pheromones are added on RReq route. Then if a RRep crossed the same route, more pheromones are added. Route is then selected according to pheromones density. Furthermore, pheromone evaporates with time and are added by transmitted packet to maintain and modify route if needed. MAR-DYMO also uses a Kinetic Graph framework to make prediction about node's neighbors trajectory. It uses aperiodic HELLO message sending compared to DYMO and reduce Bandwidth consumption.

### 2.4.1.9 QoSBeeVANET

Quality of Service Bee Swarm routing protocol (QoSBeeVANET) is a topology based protocol designed for unicast routing inspired over bee swarm. In this protocol, the first phase which is route request (RReq) is implemented using stochastic broadcasting. Which means, every node of the network is given a random or determined probability to forward a message. This type of broadcasting helps reducing the number of bandwidth consumption. As soon as the destination is reached a RRep is sent back to the source and the route is stored in a routing table with the following information: next hop, number of hop before destination, hop count. The algorithm maintains routes by sending periodically packets to neighbors and if it detects a missing node or a loss of Quality of Service (such a too high bandwidth consumption or end to end delay), it will warn all other nodes concerned on the degraded path. All nodes on error are removed and a new route discovery phase is started. After all, this algorithm can easily flood the network because of the number of packets send, especially if an error occurs (due to node missing or QoS not respected).

### 2.4.1.10 HyBR

Hybrid Bee Swarm Routing (HyBR) [23] has been designed to overcome drawbacks faced by QoSBeeVANET. Still designed to propose unicast and multicast routing, HyBR use two types of routing depending on the density of the network. Topology based routing when the density is high and geography-based routing when the density is low. The topology based routing RReq and RRep is executed like in QoSBeeVANET using stochastic flooding for RReq. The RRep is then routed back to destination throughout discovered path and stored in a table.

The geography-based routing is based on shortest node algorithm. A broadcasting flooding is executed to determine all routes to destination. Then the route is selected hop by hop with the node the closest in hop distance to the destination. Results obtained by this algorithm are close to those obtained with AODV and GPSR.

### 2.4.1.11 Datataxis

Datataxis is a topology based unicast routing protocol inspired by the behavior of Bio-System: Escherichia coli bacteria (an active component of in the natural immune system). Datataxis is designed to collect metadata (such as location, time stamp, etc.) in urban environment. Then those data were proposed to be diffused using the protocol MobEyes. Datataxis estimate firstly the meta-data density by road segment and then send multi-agents systems allowed to move from nodes to nodes to collect those data. This protocol has been proposed essentially for distributed surveillance and monitoring, for police car for example. Therefore, we will not detail it in more detail.

### 2.4.1.12 MURU

Multi-hop routing for Urban VANET is a topology based routing protocol designed for unicast and multicast communication. The protocol is based on AODV but instead of using hop count to find optimal path to destination, the Expected Disconnection Degree (EDD) is used. EDD is calculated over the Packet Error Rate (PER) of link. PER being function of hop distance, EDD being the probability that a link break then depends mainly on hop distance. EDD depends also independently to predicted speed, movement trajectory and vehicle location. Knowing that in the paper, the vehicle's mobility is approximated to a first order Markov chain. The Markov chain define a stochastic process in which the conditional probability distribution of future state depends only on the current state.Besides this improvement, the first phase of the protocol, the route request broadcasting is constrained by vehicle movement trajectory. This protocol shows good result but the number of information required to calculate EDD can be difficult to obtain in real life scenario. Furthermore, as other topology protocol, maintaining path topology decrease scalability of the protocol.

### 2.4.2 Position Based Routing

The lack of scalability and robustness of topology-based protocol has lead research to find other type of routing protocols. Indeed, the creation and maintaining of routing table in highly scalable networks may not be reliable. Position-based routing protocols which use network location of nodes to decide how to route messages are a new area of research. Although, Geocast routing can be defined as a position based routing protocol because it defines area based on geographical coordinates where nodes are concerned about the message.

### 2.4.2.1 GPSR

Greedy Perimeter Stateless Protocol (GPSR) [26] uses periodic beacon messages to build neighbors table on each node. The next-hop selection is distance based and uses GPS node's location. GPSR also integrates recovery strategies based on perimeter routing to eliminate redundant routes. This protocol is the most used in VANET to run simulation because it presents good reachability and end-to-end delay.

### 2.4.2.2 AMAR

Adaptive Movement Routing (AMAR) like GPSR, uses a greedy forwarding technique for next hop selection. But in addition to the distance, AMAR also use neighbors' position, direction and speed to select the next hop.

### 2.4.2.3 GYTAR

Like precedent protocol, Greedy Traffic Aware Routing (GYTAR) bases its next hop selection on greedy forwarding. GYTAR also use periodic beacon to maintain a table of neighbors containing position, velocity and direction. Secondly, it defines junction based of nodes density close to the node. GYTAR uses then the table and the junction density to select next hop between its neighbors.

### 2.4.2.4 DREAM

Distance Routing Effect Algorithm Mobility (DREAM) acquires each node's position using local services. It then calculates the possible destination area position and use directional flooding to reach it. The directional flooding consists of restricting the flooding graph to nodes in the area that leads to the destination.

### 2.4.2.5 LABAR

Location Area Based Ad-hoc Routing (LABAR) [27] uses a backbone next hop selection using V2I communication (with road side unit) to create an infrastructure in the network. LABAR then routes message from mobiles nodes using fixed backbone nodes. To determine the route among fix nodes it uses directional routing such as AODV.

### 2.4.2.6 ROVER

Robust Vehicular Routing (ROVER) [28] represents an example of Geocast routing. Geocast routing consider that only certain area is concerned by the message sent. Those area are called Zone Of Relevance (ZOR) and are defined by their GPS location. Each packet is then affected to one ZOR and will be forwarded to each node in this area. Each node discovers in which ZOR it belongs using local services.

### 2.4.2.7 pPSO (AODV)

The parallel Particle Swarm Optimization (pPSO) for VANET is inspired by swarm and is a parametrization of the protocol AODV. It consists of calculating the best position and speed for the vehicles to occupy in order to make the protocol AODV reduce its packet overhead, end-to-end delay and delivery ratio.

### 2.4.2.8 GSR

Geographic Source Routing (GSR) uses a Reactive Location Service (RLS) and a digital map to handle routing in urban area. With the location of the destination acquired with RLS and fixed node in the network (RSU at intersections), GSR use Dijkstra to calculate the shortest path between junction (fixed nodes) and greedy forwarding to disseminate the packet.

### 2.4.2.9 CAR

Connectivity-Aware Routing (CAR) uses neighbor recognition using beacon messages sent with a time interval depending on the number of neighbors detected. The route request phase use anchor points selected over best quality link method. Then the packet is forwarded using a greedy forwarding method among those anchor points.

### 2.4.3 Cluster Based Routing

Cluster based routing consists of dividing the network into smaller groups of nodes. Among each group a cluster head will be selected, basically a node will handle every communication. The ones inside the cluster but also the one outside it. The main objective of cluster based routing is to handle high scalability of VANET by handling smaller connected infrastructure networks. Although, this type of protocol often uses GPS information to delimit and organize its cluster.

### 2.4.3.1 CBLR

The Cluster Based Location Protocol (CBLR) builds its cluster using bacon hello messages as an initialization phase. This operation is also used to define the cluster head that will construct and maintain a table containing information over nodes in the cluster and others cluster heads of the network. Dissemination inside the cluster is effected using a greedy forwarding techniques. Communication outside of the cluster is handled firstly by finding the location of the destination using other cluster head information

### 2.4.3.2 CBDRP

Cluster Based Directional Routing Protocol (CBDRP) builds its cluster on nodes velocity vector (speed and direction). The routing is then effected like in the CBLR protocol.

### 2.4.3.3 EDCBRP

Euclidian Distance Cluster Routing Protocol (EDCBRP) bases the network division in cluster on Euclidian distance. Nodes with a Euclidian distance with each other under a defined threshold form a group. The distance is calculated using GPS information of nodes and acquired by hello message beacon periodically send. Topology tables are maintained inside clusters. For communication with other cluster, a route-request route response is sent in order to build the route to destination.

### 2.4.3.4 TACR

Trust Dependent Ant Colony Routing (TACR) is a Bio inspired routing algorithm for VANET. In TACR, cluster are builds on position and speed of moving nodes in the network. The cluster head is selected on lowest node speed with priority to RSU because of their fixed position and infrastructure network available. The communication between cluster is achieve using the Ant Colony Optimization algorithm. It consists of a route request sent to every other cluster that check if the destination is in its member table. If yes, the cluster head do not forward the route request but instead, send a route response backward. Inside cluster communication are maintained with maintaining a member table using Beacon messages.

### 2.4.4 Trajectory Based Routing

Trajectory based routing are developed mainly for urban environment with Road Side unit (RSU) positioned on roads intersection. Those routing protocols propose to use a fixed infrastructure composed by the RSUs and disseminate message to moving node using trajectory calculation. This type of algorithm modifies the V2I and I2V communication by adding several information transmitted to and by fixed RSU. Thereby, traffic statistics such as density, average speed, average direction or digital map of the area can be transmitted over V2I.

### 2.4.4.1 AMR

The Adaptive Message Routing (AMR) algorithm has for main objective to reduce the end-to-end delay. To achieve its goal, it builds route using a genetic algorithm based on the probability of connectivity and the hop count. AMR uses backbone next hop selection, using fixed RSU to convey messages over the network. The algorithm calculates the intersection between the source and the destination and the infrastructure network build on RSU.

### 2.4.4.2 IGRP

The Intersection-based Geographical Routing Protocol (IGRP) is mostly used to send packet from vehicles to the internet using a genetic optimization algorithm over intersection routing protocol. This algorithm uses a backbone next hop selection technique among RSUs until it reaches an internet point.

### 2.4.4.3 TBD

The Trajectory Based Data (TBD) algorithm uses vehicular density, mobility patterns, average speeds and digital map transmitted over V2I communication to evaluate its best next-hop to reach the closest RSU with the lower end-to-end delay. Then, it shares this delay with close nodes for them to build their own path to RSU. Packets are sent over RSU network that will disseminate it to destination using classical infrastructure networking routing.

### 2.4.4.4 TSF

Trajectory-based Statistical Forwarding (TSF) calculate end-to-end delay in the opposite way of TBD, meaning from the fixed node to the moving vehicles. Then based on the minimal delay between nodes and RSU, the route between source and destination is calculated with destination trajectory. An optimal target point is identified between the destination node's trajectory and an RSU on the network. Therefore, the packet will be forwarded over the RSU network to reach the optimal target point at the same moment as the destination node.

### 2.4.4.5 TMS

Trajectory-based Multi-Anycast (TMS) assumes the existence of a Traffic Control Center (TCC) containing information of all nodes in the network (position and velocity). Based on those information, every time a packet is sent, the TCC identify a rendezvous point between destination node's trajectory and a forwarding tree build on moving nodes.

### 2.4.4.6 STDFS

Shared-Trajectory-based Data Forwarding Scheme (STDFS) uses RSU to propagate nodes' trajectory over the network. With those information, every node can calculate a rendezvous point with the destination and disseminate the packet over V2V communication to the target point.

### 2.4.5 Geographic and Broadcast Routing

Broadcast routing protocol are mainly used to transmit warning information or other data that concern every vehicle on the road. However, this type of routing is also used as the route request phase for some multi-cast or unicast protocols. For the route discovery phase for example. The main goal of such routing is to reduce bandwidth consumption by skipping the route discovery phase and therefore, decreasing the number of routing packet sent on the network.

### 2.4.5.1 Hybrid-DTN

Hybrid geographic and Delay Tolerant Networks (Hybrid DTN) uses a greedy forwarding with a furthest node next hop selection as a route discovery phase. However, if the greedy forwarding fail, it uses the perimeter forwarding (or right-hand rule) method to reach the destination. This method consists for a node, of covering a counterclockwise circle around itself and forwarding the packet to its first neighbor found in this circle.

### 2.4.5.2 SRB

Secure Ring Broadcasting (SRB) is based on the best quality link selection. Indeed, it classify nodes based on receiving power to estimate the distance between the receiver node and the last broadcasting one. Then only those at the preferred distance from the last broadcasting node can forward the packet several times. This algorithm is based on a flooding techniques. However, depending on the estimated distance between graph level, nodes can forward one or several times.

### 2.4.5.3 PBSM

Parameter less Broadcast in Static to highly Mobile (PBSM) only checks if nodes in the neighborhood has received the packet and then retransmit the packet only to those who did not receive the packet.

### 2.4.5.4 EAEP

Edge-Aware Epidemic Protocol (EAEP) uses probabilistic techniques to decide whether to forward the packet or not. A time probability is calculated by each node to decide when to forward the packet.

### 2.4.5.5 DV-CAST

Distributed Vehicular broadcast (DVCAST) uses periodic beacon to maintain a table of neighbors for each node to know local connectivity. Then, depending on the connectivity of each node, action of forwarding is decided. DV-CAST sort the node in two categories, the well-connected ones and the sparely connected ones. The next forwarding node in the well-connected one is selected over the distance with the sender node. Indeed, a back off timer is calculated inversely proportional to the distance with the sender. The node with the smaller back off time (the furthest node) will then rebroadcast the message. For the second category, the sparely connected ones, if a node has a neighbor in the opposite direction road it will rebroadcast immediately, if not, then it will keep the packet until it finds another node in the opposite direction road.

### 2.4.5.6 TRADE

TRAck DEtection (TRADE) protocol gives nodes a table of neighbors maintained using periodic beaconing. A node sorts its neighbors in three categories depending on their position and velocity: same road ahead, same road behind and different road. Then it uses the furthest node selection technique as next hop selection in the two first categories. For the third one, the node just rebroadcast to every node in this category.

### 2.4.5.7 MAC

The Media Access Control (MAC) protocol also uses periodic beaconing to maintain a table of neighbors on each node. Consequently, every node calculates its relative direction to the sender and the ones between the sender and its neighbors. Then if the node's relative direction corresponds to the packet direction, the node will check in its neighbor table if it is the more dedicated to forward. This selection is based on the furthest node criteria using the segment define by the distance and the relative direction calculated before.

### 2.4.5.8 REAR

Reception Estimation Alarm Routing (REAR) protocol is based on probabilistic next hop selection among neighbors in the direction of the message propagation. REAR makes node maintain a neighbor table and each warning message sent contains sender position and a list of sender's neighbors. Each node in the direction of the message's direction will then calculate its reception probability and wait a back off time inversely proportional to this value. The first node to reach its back off time will then rebroadcast the message.

### 2.4.5.9 GPCR

Greedy Perimeter Coordinator Routing (GPCR) defines junction as a link between one node reached by a message and another one out of the message's range. It uses 2-hop neighbors table to locate a junction between to nodes. Indeed, by sending periodic beacon containing its position and its neighbors, each node can construct such a table. Therefore, if a node has two neighbors that does not have each other on their respective table, that means this node is a junction. Every node calculates the number of junction it represents and the those that have the bigger coefficient are called coordinator. Coordinators informs other nodes of their new role and form the backbone of the network. Consequently, the next forwarding hop selection during the broadcasting will prefer a backbone node as the next forwarding node. If no coordinator is found, the furthest node method is used.

### 2.4.5.10 TDR

Three-Dimensional scenario oriented Routing (TDR)  protocol proposes an improvement of GPCR protocol and its upgrade GyTAR to fit 3-Dimensional environment. The main difference comes from the size of hello beacons which contain 3 coordinates instead of two.

### 2.4.5.11 Multicast Routing for Message dissemination protocol

This protocol also uses beaconing to maintain neighbor's awareness table and the next hop selection is based on the most demanding node criteria.

### 2.4.5.12 OAPB

Optimistic Adaptive Probabilistic Broadcast (OAPB) constructs and maintains for each node a 2-hop neighbors table using periodic beacon sending. Then each node calculates a probability of rebroadcasting based on that information. it create a back off timer, function of the probability of rebroadcasting and a random variable. The node with the smallest back off timer will be selected as the next hop for the message.

### 2.4.5.13 UMB

Urban Multi-Hop Broadcast (UMB) protocol considers that every message should be transmitted to RSU disposed to every intersection and that should behave as repeater. If there is any RSU available, the algorithm will function on a reactive way. When message needs to be sent it will send a directional request to relay (RTS) containing its position and direction of propagation. The nodes in the broadcast area defined by the packet direction emit a signal of a duration proportional to the distance with the sender and the number of jamming signal in its transmission range. The node with the longest jamming (i.e. the furthest away from the sender) signal will send the Clear To Broadcast answer (CTB) to the sender and will be designed as next hop. This method can be classified as furthest node selection. When the warning message is sent, elected relay will send back an ACK message to the sender. If it fails to do so or if several CTB are received, UMB will start the recovery algorithm.

### 2.4.5.14 SB

Smart Broadcast (SB) is based on UMB but replaces the jamming signal of the furthest node selection method by a classic back off timer inversely proportional to the distance with the transmitter. This way the next relay is not the one that wait the most of time like in UMB. SB also handle CTB collision better with a random selection between the two possible routes.

### 2.4.5.15 STAR

Intersection based routing protocol that uses RSU placed at red light intersections. Packet are forwarded on every red light connected and to every first car on green light that are the closest to the destination.

### 2.4.5.16 VanetDFCN

VANET Delayed Flooding with Cumulative Neighborhood (VANETDFCN) is an improvement of the protocol DFCN for MANET. It transmit more information about neighbors (such as the position) to improve the forwarding node selection method. The next hop selection is distance based, if the distance between the receiver and the sender is over a Distance To Live, the node will not forward. The packet must also have been received only once. Based on those criteria, a coefficient is calculated which represent the number of TCP chunk a packet can be divided in and transmitted on the liaison created.

### 2.4.5.17 xChangeMobile

xChangeMobile is an improvement of VanetDFCN with the addition of a Threshold on the minimal number of chunk that can be transmitted over a liaison. Only node that can provide a liaison with a coefficient higher than the Threshold will be considered as forwarding node.

### 2.4.5.18 MOCell

MOCell routing protocol is based on a genetic algorithm to improve xChangeMobile. The goal is to reduce bandwidth consumption and the number of lost chunks in the TCP process.

### 2.4.5.19 RBLSM

Reliable Broadcasting of Life Safety Messages (RBLSM) is also a reactive protocol that sends RTS and waits for CTB when a packet needs to be sent. However, RBLSM uses the nearest node next hop selection instead of the furthest.

### 2.4.5.20 LW-RBMD

Light Weight Reliable Broadcast Message Delivery (LW-RBMD) protocol uses the furthest node technique using only the header of warning messages to transmit sender's position. Nonetheless, the sending node will wait for a rebroadcasted message and take it into an acknowledgment (ACK). If the ACK is not received, the sender will resend the packet after a timer. This protocol is designed to limit the network overhead and still maintaining high reliability.

### 2.4.5.21 MHVB

Multi-Hop Vector Broadcasting (MHVB) is a broadcasting routing algorithm using the furthest node selection technique with a classical inversely proportional back off timer set up. However, MHVB integrates a congestion detection algorithm, which consist of detecting when the number of neighbors is above a certain Threshold. In that case, the Broadcasting interval is increased to lower the bandwidth consumption.

### 2.4.5.22 DTM

Distance-To-Mean (DTM) algorithm uses the distance to mean as next hop selection technique. Each node maintains a table of neighbors using beaconing and defines a distance threshold based on the number of neighbors. This distance represents the minimum the distance to mean value must exceed so that the node is considered as a possible forwarding node. Firstly, introduced in 2011, it presents better results than distance based greedy forwarding in terms of reachability and of network overhead. This paper being based on this paper, DTM will be explained in more details in the next section.

### 2.4.5.23 DADCQ

Distribution-Adaptive Distance with Channel Quality (DADCQ) protocol uses both distance, and best quality link as next hop selection. Here, warning messages also propagate their neighbor table in their header. This way, each node has access to two-hop neighborhood.

### 2.4.5.24 CSBD

CSBD  is a MAC and network cross layer with density-adaptive contention window. This means that the MAC layer is directly influenced by the information obtained on the network layer using geographical routing and distance-to-mean heuristic. [51].

### 2.4.5.25 SLAB

Statistical Location Assisted Broadcast (SLAB) enhances DADCQ using the distance-to-mean next-hop selection method instead of the distance-based used in DADCQ. Also, SLAB uses machine learning techniques to improve the Threshold function definition.

### 2.4.5.26 FLB

The Fuzzy Logic Based (FLB) protocol is based on the DTM algorithm but uses fuzzy-based techniques to calculate the threshold value. Indeed, each node is classified according to several factors defined on its information such as velocity, position and number of neighbors; the Coverage factor, the Mobility Factor and the Connectivity Factor. This improvement in the broadcasting node selection presents better results than DTM in terms of number of rebroadcast per covered nodes.

### 2.4.5.27 BEFLAB

Bandwidth Efficient Fuzzy Logic-Assisted Broadcast (BEFLAB) for VANET, like FLB presents a Fuzzy-logic based receiver based rebroadcasting node selection. Each node is here classified according to two factors, the Mobility Factor and the Coverage factor. Then a table of Fuzzy rules decide whether a node will rebroadcast or not. Unlike FLB, BEFLAB does not use the distance-to-mean heuristic method but only the fuzzy technique.

### 2.4.5.28 IHAB

Intelligent Hybrid Adaptive Broadcast (IHAB) for VANET is based on FLB and BEFLAB. Indeed, IHAB first calculate the node potential transmit density (PTD) using two-hop neighbor's information. Then IHAB chooses between FLB and BEFLAB which protocol to use. If the network is said dense (PTD superior to a threshold), IHAB will chose to use BEFLAB, if the network is sparse (PTD inferior to threshold), IHAB will chose to use FLB.

# CHAPTER 4

## METHODOLOGY

Simulations have been carried out on NS3 to compare and analyze routing algorithms, such as the DSDV, AODV, and DSR, based on various performance metrics. However, performance comparison and analysis between the two classical MANET routing protocol types, proactive and reactive, have rarely been done using NS3 in the Linux Ubuntu operating system. The procedures and simulation results presented in this project report will help VANET researchers and designers tune their systems to meet particular requirements in a more efficient way.

**NS3 Simulation Modeling Methodology**

To establish NS3 simulations, several classes such as core-module.h and network-module.h need to be included. These classes plus their detailed descriptions can be found in NS3 API. Moreover, NS3 employs C++ and Python languages, and several simulation steps need to be followed to start any NS3 simulations.

**3.1.1 Simulation Parameter Setup.**

| Simulation Parameter Setup | |
|---|---|
| Description | **Parameters** |
| Channel/Wireless channel | Channnel Type |
| Propagation | Radio-Propagation Model |
| Phy/Wireless Phy | Network Interface Type |
| Mac/802.11 | MAC Type |
| Queue/Droptail/Priqueue | Interface Queue Type |
| LL | Link Layer Type |
| Antenna/OmniAntenna | Antenna Model |
| 50 | Max Packet |
| 100 | Number Of Mobile Nodes |
| AODV,DSR,DSDV | Routing Protocol |
| 867 | Topographical Dimention |
| 561 | Topographical Dimention |
| 25s,50s,75s,100s,125s,150s,175s,200s. | Time Of Simulation |

Table-2

# CODE   SECTION

## CODE SNIPPETS

### 1.MESH.

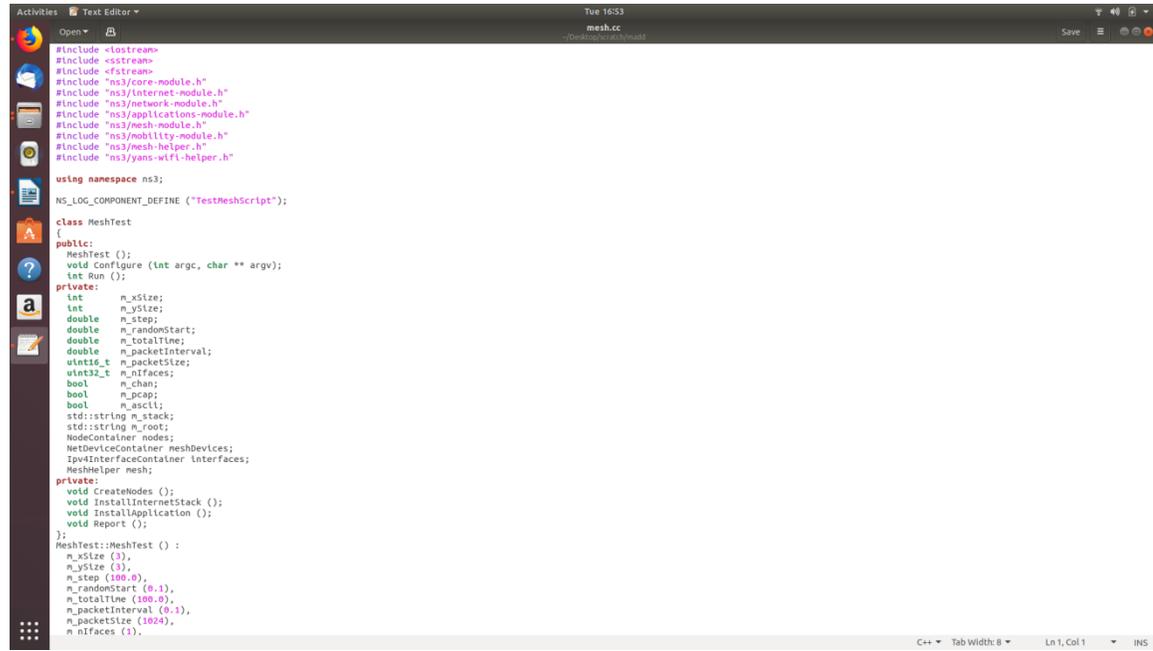

### 2.AODV

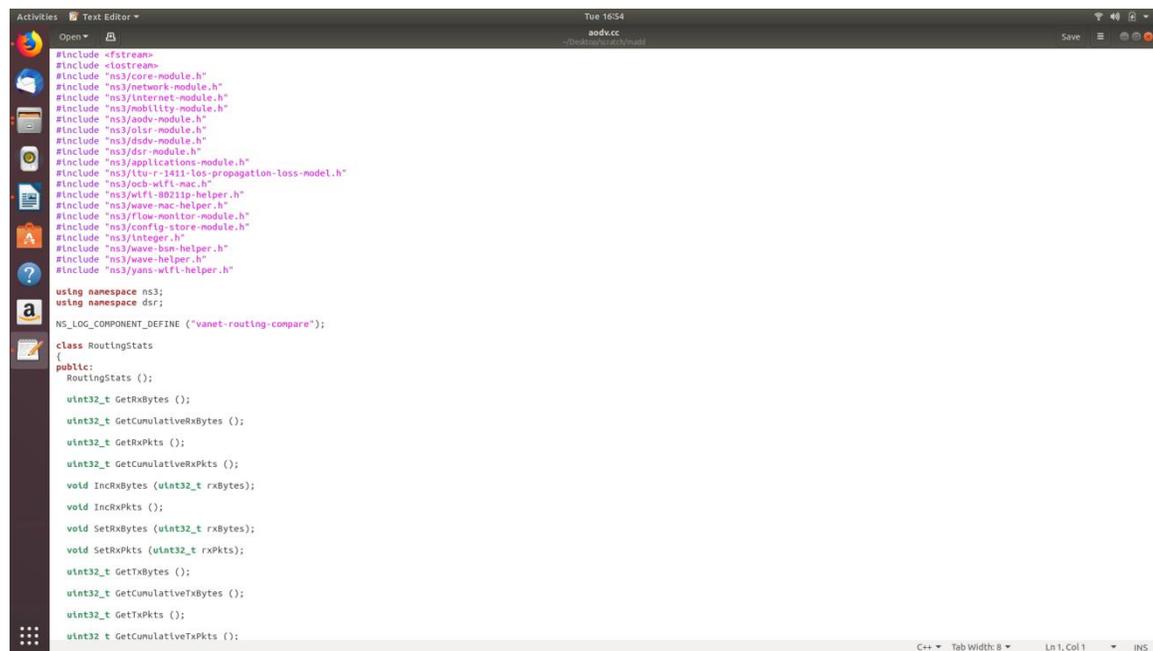

## 3. DSR

## 4. DSDV

## 4.1 DSDV ROUTING TABLE

```
#ifndef DSDV_RTABLE_H
#define DSDV_RTABLE_H

#include <cassert>
#include <map>
#include <sys/types.h>
#include "ns3/ipv4.h"
#include "ns3/ipv4-route.h"
#include "ns3/timer.h"
#include "ns3/net-device.h"
#include "ns3/output-stream-wrapper.h"

namespace ns3 {
namespace dsdv {
enum RouteFlags
{
  VALID = 0,     // !< VALID
  INVALID = 1,   // !< INVALID
};

class RoutingTableEntry
{
public:
  RoutingTableEntry (Ptr<NetDevice> dev = 0, Ipv4Address dst = Ipv4Address (), uint32_t seqNo = 0,
                     Ipv4InterfaceAddress iface = Ipv4InterfaceAddress (), uint32_t hops = 0, Ipv4Address nextHop = Ipv4Address (),
                     Time lifetime = Simulator::Now (), Time SettlingTime = Simulator::Now (), bool changedEntries = false);

  ~RoutingTableEntry ();
  Ipv4Address
  GetDestination () const
  {
    return m_ipv4Route->GetDestination ();
  }
  Ptr<Ipv4Route>
  GetRoute () const
  {
    return m_ipv4Route;
  }
  void
  SetRoute (Ptr<Ipv4Route> route)
  {
    m_ipv4Route = route;
  }
  void
  SetNextHop (Ipv4Address nextHop)
  {
    m_ipv4Route->SetGateway (nextHop);
  }
  Ipv4Address
  GetNextHop () const
  {
    return m_ipv4Route->GetGateway ();
  }
  void
  SetOutputDevice (Ptr<NetDevice> device)
```

## 5. SUMO

## 5.1 OSM MAP OF SRINAGAR CITY.

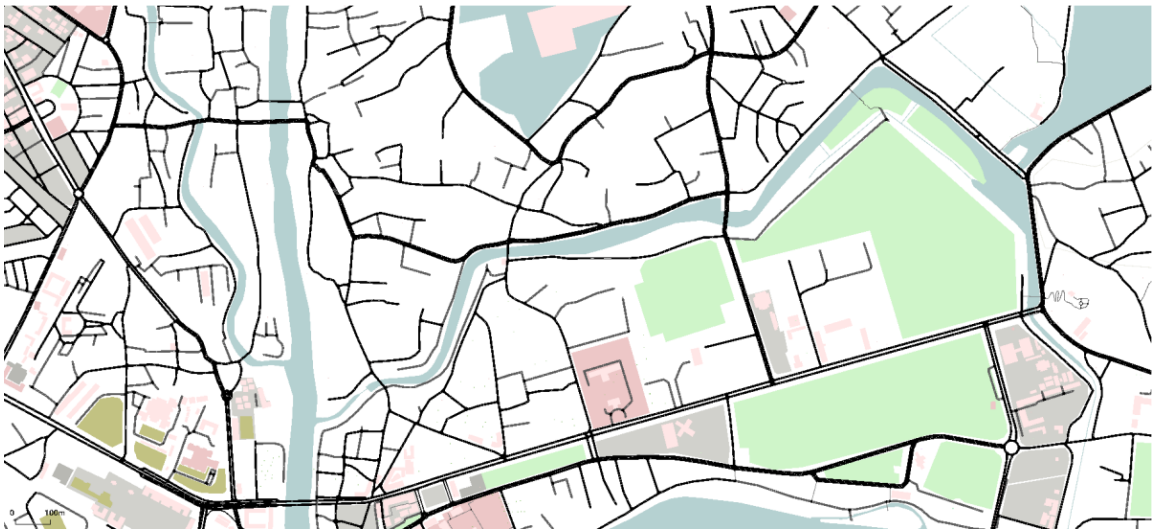

## 5.2 POLY FILE CODE

```xml
<polygonTypes>
    <polygonType id="waterway"            name="water"       color=".71,.82,.82" layer="-4"/>
    <polygonType id="natural"             name="natural"     color=".55,.77,.42" layer="-4"/>
    <polygonType id="natural.water"       name="water"       color=".71,.82,.82" layer="-4"/>
    <polygonType id="natural.wetland"     name="water"       color=".71,.82,.82" layer="-4"/>
    <polygonType id="natural.wood"        name="forest"      color=".55,.77,.42" layer="-4"/>
    <polygonType id="natural.land"        name="land"        color=".98,.87,.46" layer="-4"/>

    <polygonType id="landuse"             name="landuse"     color=".76,.76,.51" layer="-3"/>
    <polygonType id="landuse.forest"      name="forest"      color=".55,.77,.42" layer="-3"/>
    <polygonType id="landuse.park"        name="park"        color=".81,.96,.79" layer="-3"/>
    <polygonType id="landuse.residential" name="residential" color=".92,.92,.89" layer="-3"/>
    <polygonType id="landuse.commercial"  name="commercial"  color=".82,.82,.88" layer="-3"/>
    <polygonType id="landuse.industrial"  name="industrial"  color=".82,.82,.88" layer="-3"/>
    <polygonType id="landuse.military"    name="military"    color=".60,.60,.36" layer="-3"/>
    <polygonType id="landuse.farm"        name="farm"        color=".95,.95,.80" layer="-3"/>
    <polygonType id="landuse.greenfield"  name="farm"        color=".95,.95,.80" layer="-3"/>
    <polygonType id="landuse.village_green" name="farm"      color=".95,.95,.80" layer="-3"/>

    <polygonType id="tourism"             name="tourism"     color=".81,.96,.79" layer="-2"/>
    <polygonType id="military"            name="military"    color=".60,.60,.36" layer="-2"/>
    <polygonType id="sport"               name="sport"       color=".31,.90,.49" layer="-2"/>
    <polygonType id="leisure"             name="leisure"     color=".81,.96,.79" layer="-2"/>
    <polygonType id="leisure.park"        name="tourism"     color=".81,.96,.79" layer="-2"/>
    <polygonType id="aeroway"             name="aeroway"     color=".50,.50,.50" layer="-2"/>
    <polygonType id="aerialway"           name="aerialway"   color=".20,.20,.20" layer="-2"/>
    <polygonType id="highway.services"    name="services"    color=".93,.70,1.0" layer="-2"/>

    <polygonType id="shop"                name="shop"        color=".93,.70,1.0" layer="-1"/>
    <polygonType id="historic"            name="historic"    color=".50,1.0,.50" layer="-1"/>
    <polygonType id="man_made"            name="man_made"    color="1.0,.90,.90" layer="-1"/>
    <polygonType id="man_made.pipeline"   name="pipeline"    color="1.0,.90,.90" layer="-1"/>
    <polygonType id="building"            name="building"    color="1.0,.90,.90" layer="-1"/>
    <polygonType id="amenity"             name="amenity"     color=".93,.70,1.0" layer="-1"/>
    <polygonType id="amenity.parking"     name="parking"     color=".72,.72,.70" layer="-1"/>
    <polygonType id="power"               name="power"       color=".10,.10,.10" layer="-1" discard="true"/>
    <polygonType id="highway"             name="highway"     color=".10,.10,.10" layer="-1" discard="true"/>

    <polygonType id="boundary"            name="boundary"    color="1.0,.33,.33" layer="0" fill="false" discard="true"/>
    <polygonType id="admin_level"         name="admin_level" color="1.0,.33,.33" layer="0" fill="false" discard="true"/>
    <polygonType id="place"               name="admin_level" color="1.0,.5,.0"   layer="0" fill="false" discard="true"/>

</polygonTypes>
```

# CHAPTER 5

# RESULTS & COMPARISON.

## 1. MESH RESULTS

## 2. DSDV RESULT

## 3. AODV

## 4. DSR

# COMPARISON

## 1. SIMULATION DATA FOR DSDV.

# Simulation Data For DSDV

| Simulation Time | Packet Sent | Packet Recieved | PDR | Packet Forwarding |
|---|---|---|---|---|
| 25 | 12208 | 125 | 7145 | 872 |
| 50 | 24416 | 397 | 6150.13 | 1779 |
| 75 | 36622 | 768 | 4768.49 | 5585 |
| 100 | 48830 | 1536 | 3179.04 | 9004 |
| 125 | 61063 | 2501 | 2440.46 | 12241 |
| 150 | 73244 | 3485 | 2101.69 | 15439 |
| 175 | 85450 | 4520 | 1890.49 | 18606 |

## 2. SIMULATION DATA FOR DSR.

# Simulation Data For DSR

| Simulation Time | Packet Sent | Packet Recieved | PDR | Packet Forwarding |
|---|---|---|---|---|
| 25 | 12208 | 976 | 1250.82 | 3233 |
| 50 | 24416 | 1916 | 1274.32 | 6497 |
| 75 | 36622 | 2835 | 1291.78 | 9783 |
| 100 | 48830 | 3765 | 1296.95 | 13059 |
| 125 | 61036 | 4693 | 1300.58 | 16336 |
| 150 | 73224 | 5618 | 1303.74 | 19621 |
| 175 | 85450 | 6550 | 1304.58 | 22898 |

**Table – 4**

**3.SIMULATION DATA FOR AODV.**

# AODV Simulation Data

| Simulation Time | Packet Sent | Packet Recieved | PDR | Packet Forwarding |
|---|---|---|---|---|
| 25 | 12208 | 713 | 1712.70 | 3506 |
| 50 | 24416 | 1431 | 1706.22 | 7007 |
| 75 | 36622 | 2156 | 1698.61 | 10501 |
| 100 | 48830 | 2889 | 1690.20 | 13990 |
| 125 | 61063 | 3618 | 1687.01 | 17473 |
| 150 | 73244 | 4363 | 1678.75 | 20948 |
| 175 | 85450 | 5106 | 1672.85 | 24427 |

**4. SIMULATION (AODV & DSDV) vs PDR(Packet Dessimation Rate).**

# Simulation(AODV & DSDV) vs PDR

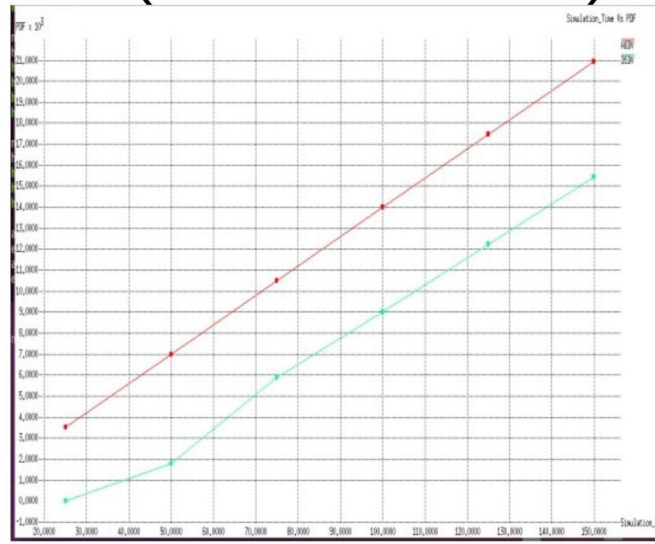

**5. AODV vs DSDV THROUGHPUT**

## Simulation Time vs Throughput

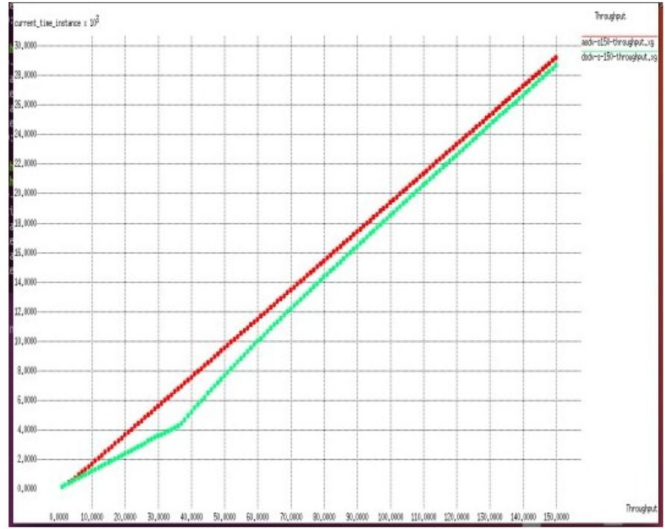

**6. AODV vs DSDV vs DSR THROUGHPUT.**

## Average Throughput

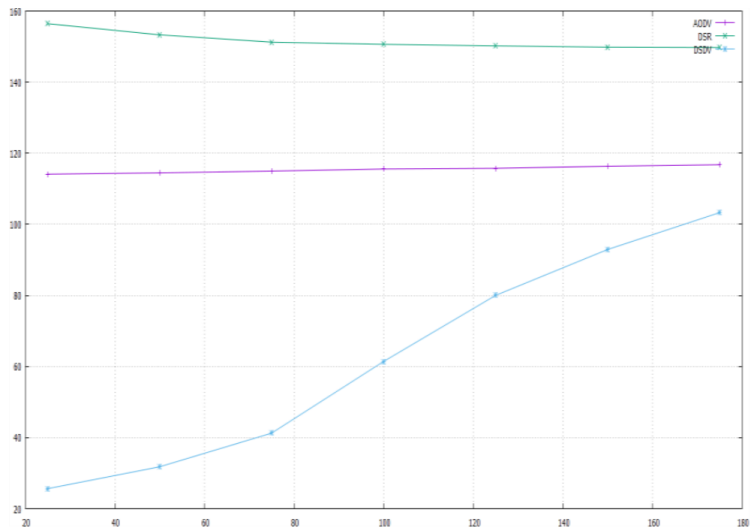

The above grahps and results depict that initaially with lesser number of nodes all the protocols perform same with a slight edge to DSR protocol on the routes of Srinagar city but later as the number of nodes increase AODV shows the greater output as compared to others

Average Throughput for lesser number of nodes: DSR > AODV > DSDV.

Average Throughput for increased number of nodes: AODV > DSDV > DSR.

Since the number of nodes is large so second scenario fits the best for routing in Srinagar city.

PDR is higher for AODV is good for lesser number of nodes but it falls down as the nodes increase and DSR takes up.

PDR for lesser number of nodes: AODV > DSR > DSDV

PDR for increased number of nodes: DSR > AODV > DSDV.

End to end delay of DSR is least and that of DSDV is highest for lesser number of nodes but when the number of nodes is increased the trend changes e2e delay gets decreased for DSDV and AODV but increases for DSR.

E2e delay for lesser number of nodes: DSR < AODV < DSDV.

E2e delay for increased number of nodes: DSDV < AODV < DSR.

Normalized Routing Load is initially zero but increases gradually with time giving less values for DSDV than AODV and DSR.

Initial NRL is equal to zero for all the three protocols.

NRL after some time: DSDV < AODV < DSR.

So from above comparison we can depict that AODV is having highest Throughput and is average in other cases of comparison for the routes of srinagar, So based on above comparison I would preffer AODV over DSDV and DSR.

# CHAPTER 6

# CONCLUSION AND FUTURE WORK

A) Average Throughput Throughput indicate rate of communication per unit time. Throughput in this experiment evaluated for AODV, DSDV and DSR for all these three mobility models. The throughput (bytes per simulation time ) versus increasing number of nodes of protocols by using column mobility of nodes in given environment. In this AODV perform better than other protocols, but in lesser number of nodes performance of all protocols is almost same. Throughput of protocols with respect increasing nodes shows that throughput of AODV is better than DSDV, and DSR perform least. Throughput by using random mobility model is much better than other two mobility models. In this case all three protocols perform better but AODV is much better.

**B) Packet delivery ratio**

It is the ratio of data packets delivered to the destination to those generated by the source. It is calculated by dividing the number of packet received by the destination through the number of packet originated by the source. The packet delivery ratio of AODV is good for lesser number of nodes but when nodes increase then PDR fall down. DSR is having increased PDR than AODV and DSDV for greater number of packets.

**C) End to end delay**

It is the amount of time taken by packet to reach from one node to other. End to end delay versus increasing number of nodes by using column mobility model. At lesser number of nodes the e2e delay of DSR is at its least value but DSDV at its peak value, after increasing number of nodes e2e of DSR start increasing but oppositely DSDV and AODV start decreasing. DSR have higher e2e delay but DSDV have least. DSR is on its lowest value and DSDV have its largest point. Generally, With the increasing number of nodes the delay of these protocols gradually decrease.

**D) Normalized routing load**

It is the metadata and network routing information sent by a source node to destination node, which uses a portion the available bandwidth of a protocol. AODV have a higher NRL but DSDV is on bottom. Initially value of NRL is zero, but with number of nodes it's gradually starts increasing. Value of NLR in DSDV is very lesser. In case of group mobility NRL is high when routing protocol is DSR and less when DSDV. If random way point mobility model is used it shows a less values for DSDV but high value for DSR and AODV.

Srinagar city needs implementation of these protocols and also intelligent traffic system with modification of some roads with better planning. Some junctions are normal with current flow of traffic while others need some modification as well some extra roads are needed.

# FUTURE WORK

1. Protocols need modification and some new protocols are needed in general.

2. Stable and more actualised simulators are needed.

3. Implementation of these protocols in our traffic system is required.

4. Better Planning is necessity.

5. Upgraded roads and intelligent vehicles.